# Multi-Sensor Methane Mapping in a Unified Framework: Tanager-1 Integration and comparison to EnMAP and PRISMA

Alvise Ferrari
*GMATICS S.r.l.,*
Rome, Italy
*School of Aerospace Engineering*
*Sapienza University*
Rome, Italy
alvise.ferrari@uniroma1.it

Valerio Pampanoni
*School of Aerospace Engineering*
*Sapienza University*
Rome, Italy
valerio.pampanoni@uniroma1.it

Giovanni Laneve
*School of Aerospace Engineering*
*Sapienza University*
Rome, Italy
giovanni.laneve@uniroma1.it

***Abstract*—Spaceborne imaging spectroscopy enables facility-scale methane ($CH_4$) plume detection and quantification by exploiting absorption structure in the ~1.65/2.3 µm windows. However, performance ultimately depends on both radiometric sensitivity and the mitigation of pushbroom artifacts such as column-dependent variability and striping. This paper reports the integration of Planet/Carbon Mapper Tanager-1 Level-1 radiances into a mature multi-sensor methane processing chain previously applied to PRISMA and EnMAP and evaluates the implications of Tanager's radiometric regime for matched-filter retrieval, plume segmentation, and IME-based flux estimation. The retrieval is based on a Clutter Matched Filter (CMF) formulation that yields methane enhancements in concentration–path-length units (ppm·m) and propagates uncertainty from radiance noise and background variability through enhancement maps, Integrated Mass Enhancement (IME), and emission rate via the IME method. Particular emphasis is placed on a column-wise CMF (CW-CMF), in which background statistics are estimated per detector column to reduce structured false positives induced by pushbroom non-uniformities. A compact radiometric comparison between PRISMA, EnMAP and Tanager-1 is performed on homogeneous high-reflectance calibration scenes to derive reference SNR spectra and striping diagnostics for all these sensors . We then demonstrate CW-CMF-only operational results on a landfill super-emitter in the Buenos Aires region, using paired Tanager-1 and EnMAP acquisitions over the same area of interest acquired on different dates . In the absence of near-simultaneous acquisitions and ground truth, results are interpreted in terms of background-limited sensitivity and uncertainty-stabilized IME/flux estimation rather than absolute accuracy.**

***Keywords— Methane emissions, hyperspectral satellites, PRISMA, EnMAP, GHGSat, EMIT, Tanager, oil and gas, landfills, remote sensing, atmospheric science, greenhouse gas monitoring, spectral analysis, emission quantification, satellite synergy, environmental monitoring.***

## I. Introduction

Methane ($CH_4$) is a major contributor to near-term climate forcing, and a substantial fraction of anthropogenic emissions originates from a limited number of large point sources. High-resolution imaging spectrometers can detect and map facility-scale plumes by resolving $CH_4$ absorption structure in the SWIR (notably the ~2.3 µm window) and applying statistical detectors such as Matched Filters (MFs) [1]–[3]. These methods transform subtle, spectrally structured absorption signatures into scalar enhancement maps that can be segmented and inverted to emission rate [4]–[6]. Prior work has demonstrated methane retrieval from PRISMA and EnMAP, including landfill and oil-and-gas applications, and has established MF approaches as the state of the art for plume mapping in complex surface scenes [7]–[9], [14], [15]. Tanager-1 extends imaging spectroscopy toward operational greenhouse-gas monitoring within the Carbon Mapper/Planet program [11]–[13]. While higher SNR regimes are generally favorable for methane retrieval [5], [6], pushbroom instruments can exhibit column-dependent radiometric variability that appears as striping in retrieved enhancement maps, producing structured false positives that complicate plume segmentation and inflate mass/flux uncertainty [11]–[13]. Therefore, sensor performance must be assessed not only through average SNR but also through the magnitude and structure of across-track non-uniformities and how these propagate through the retrieval [8].

This paper focuses on two linked objectives. First, we present a technically explicit description of an end-to-end methane processing chain applied consistently across Tanager-1, EnMAP, and PRISMA, including retrieval in concentration–path-length units (ppm·m), uncertainty propagation, and IME-based flux estimation [1]–[6]. These steps are implemented in HyGAS, a unified multi-sensor framework designed to enforce consistent assumptions across missions (retrieval, uncertainty, segmentation, IME and flux inversion) [16]. Detailed derivations, broader multi-sensor experiments, and full methodological documentation are provided in an accompanying extended journal manuscript (submitted), which serves as the primary algorithmic reference for this conference paper [17]. Second, we assess radiometric and artifact behavior in the methane window using a compact comparison based on reference SNR spectra and striping diagnostics (Figs. 1–2), and we demonstrate operational performance on a landfill super-emitter in the Buenos Aires region using paired Tanager-1/EnMAP acquisitions (Fig. 3) [8], [11], [16], [17]. Because the paired acquisitions are not near-simultaneous and no ground truth is available, we do not attempt absolute accuracy validation; instead, we quantify background-limited sensitivity, artifact mitigation, and the stability of uncertainty-aware IME/flux estimates [4], [6], [16], [17].

## II. DATA

### A. Sensors and input products

We process Tanager-1 Level-1 radiance scenes over selected AOIs, following Planet's product documentation and the Carbon Mapper system context [11]–[13]. Planet's documentation describes the Tanager imaging spectrometer (VSWIR coverage and dense spectral sampling) and highlights collection modes designed to increase sensitivity and mitigate geometric effects relevant to pushbroom imaging [12], [13]. For cross-sensor context we process EnMAP Level-1B radiances and include PRISMA in the radiometric comparison on calibration scenes [7]–[9].

A practical nuance for spatial sampling is that Carbon Mapper descriptions report that certain distributed Level-2 products are resampled to a 30 m grid, while acquisition geometry and sensitivity modes can imply non-square instantaneous sampling that is mitigated through collection strategy and product generation [11]–[13]. In this paper, enhancements are retrieved directly from Level-1 radiances for all sensors to keep a consistent methodology as implemented in HyGAS [7]–[9], [16], [17].

While all three sensors are imaging spectrometers, the retrieval sensitivity and artefact propagation depend on the combined effect of (i) radiance regime in the 2.3 μm window, (ii) effective spectral sampling / response, and (iii) pushbroom across-track non-uniformity. To isolate these factors, we process Level-1 radiances with a common workflow and apply radiance-normalised SNR diagnostics and striping indicators on homogeneous desert targets (Sec. II.B; full definitions in [17]). The analysis is intentionally focused on background-limited stability (noise floor and structured false positives) rather than absolute accuracy, consistent with the non-simultaneous revisit of the sample case study (section IV) and the lack of ground truth [4]–[6], [17].

### B. Scene selection and acquisition

SNR and striping diagnostics (Figs. 1–2) are computed on three high-reflectance, low-heterogeneity scenes (one per sensor) selected to minimise surface-driven variability in the ~2.3 μm methane window. The selected PRISMA calibration scene is acquired over Northern State (Sudan), 1 April 2020, 08:53 UTC (08:53:13–08:53:18 UTC); the EnMAP calibration scene is over Agadez (Niger), 12 July 2022, 10:43 UTC (10:43:02 UTC); the Tanager-1 is over Northern State (Sudan), 9 May 2025, 09:03 UTC (09:03:23 UTC). These three scenes are used consistently for the radiometric comparison and striping metrics reported in this paper [16], [17].

The operational demonstration focuses on the Complejo Ambiental Norte III landfill (Buenos Aires, Argentina) using one EnMAP and one Tanager-1 acquisition over the same facility acquired on different dates [17]. The two acquisitions are: EnMAP Level-1B, 9 March 2024, 14:24 UTC (14:24:31 UTC); Tanager-1 Level-1 radiance, 26 June 2025, 14:44 UTC (14:44:19 UTC).

Since the scenes are not co-temporal and do not share a common footprint, segmentation and background statistics are computed separately on each acquisition's analysis area. The comparison is therefore designed to examine background-limited noise/false positives, artefact susceptibility (striping), and uncertainty behaviour, rather than absolute accuracy [4]–[6], [16], [17].

## III. METHODS

### A. Retrieval overview and qualitative theoretical framework

Under moderate enhancements and near-nadir viewing, radiance perturbations due to $CH_4$ absorption can be treated as approximately linear in a concentration–path-length enhancement $\Delta X$ (ppm·m), motivating a MF style estimator that detects the $CH_4$ absorption template in the presence of background spectral variability (clutter) [1]–[3], [6]. This framework is consistent with established imaging spectroscopy methane retrieval literature, and the complete derivation used here is detailed in the extended manuscript [17].

Operationally, each pixel spectrum $\mathbf{x}$ is modeled as background plus a $CH_4$ absorption contribution. A methane "target" vector $\mathbf{t}$ is derived from a unit-absorption spectrum and representative background radiance, resampled into instrument space using sensor spectral response information; background statistics are characterized by mean $\boldsymbol{\mu}$ and covariance $\boldsymbol{\Sigma}$ estimated from plume-free pixels [1], [2], [6], [17]. A Clutter Matched Filter (CMF) provides $\Delta X$ by projecting the whitened residual $(x-\mu)$ onto the whitened target, yielding a scalar enhancement map [1]–[3], [6]. All processing steps described in Sections 3.1–3.4 are implemented in HyGAS [16], with methodological details consolidated in [17].

### B. Background selection and CMF variants

A key practical step is estimation of $\boldsymbol{\mu}$ and $\boldsymbol{\Sigma}$ from representative background pixels. To reduce plume contamination and avoid mixing spectrally distinct materials, background pixels are selected through a spectrally matched procedure prior to covariance estimation [6], [17]. We implement (i) a scene-wide CMF, (ii) a Cluster-Tuned CMF (CTMF) [1], [6], and (iii) a column-wise CMF (CW-CMF) where $\mu$ and $\Sigma$ are estimated per detector column (or column blocks), explicitly accommodating pushbroom non-uniformities [11]–[13], [17]. CW-CMF is used as the baseline for Tanager-1 and is the configuration reported in the case study figure in this conference paper [16], [17].

### C. Radiometric and striping diagnostics

Reference SNR spectra are estimated on the three calibration scenes listed in Section II.B using a PCA-based noise approach that separates scene structure from instrument noise, and are summarised with robust statistics across detector columns (Fig. 1) [17]. Because pushbroom imagers may exhibit column-dependent non-uniformity that propagates as structured background artefacts ("striping") in matched-filter enhancement maps, we also quantify striping on homogeneous ROIs using complementary indicators defined in the extended paper [17]. These include robust amplitude metrics (scene-based and ratio-based, the latter suppressing multiplicative surface brightness structure) and FFT-based periodicity indicators [17]. Due to space constraints we report only the ratio-based striping metric in Fig. 2, while the full multi-metric assessment and discussion are provided in [17] (see Fig. 15 therein). These diagnostics motivate the use of CW-CMF when column-dependent variability dominates enhancement-map structure, consistent with pushbroom-aware retrieval practice for Tanager products and with EnMAP methane retrieval analyses [8], [11]–[13], [16], [17].

### *D. Uncertainty propagation, IME and flux*

Enhancement uncertainty combines radiometric-noise-driven uncertainty and background-driven variability, evaluated under the same μ/Σ estimates used for retrieval [6], [17]. Plume masks are derived using a scale-aware segmentation procedure; IME is computed by integrating ΔX over plume area, and flux Q is estimated using the IME method with wind uncertainty propagated into Q [4], [6], [17].

## IV. Results

### *A. Radiance-normalised SNR comparison in the methane window (Fig. 1)*

Figure 1 reports radiance-normalised median SNR curves for PRISMA, EnMAP, and Tanager-1, derived from the calibration scenes in Section II.B and scaled to a common reference radiance regime to suppress scene-brightness effects [17]. In this paper, PRISMA and Tanager-1 $SNR_{ref}$ are explicitly scaled to the EnMAP reference radiance regime (i.e., EnMAP provides the common $SNR_{ref}$ radiance baseline). Under this normalisation, the curves isolate differences in the effective instrument noise floor relevant to methane retrieval uncertainty propagation: EnMAP exhibits a higher radiance-normalised SNR than PRISMA in the 2.3 μm methane window, and Tanager-1 indicates a further improvement, consistent with the ordering expected for instrument-noise-limited enhancement uncertainty [7]–[9], [12], [13], [17]. Importantly, the cross-track (column-dependent) variability remains a separate effect (addressed via the striping diagnostics and CW-CMF), and it can dominate structured false positives even when median SNR is high [6], [16], [17].

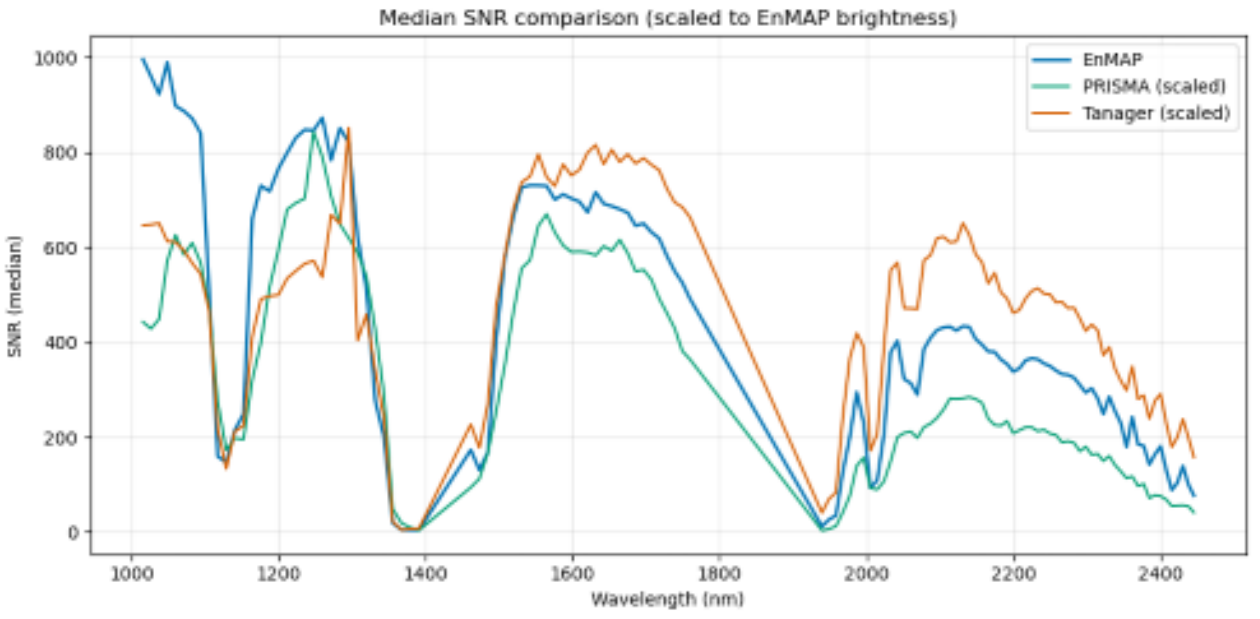

**Fig. 1.** Radiance-normalised comparison of median reference SNR spectra for PRISMA, EnMAP and Tanager-1 across the SWIR methane window, derived from the calibration scenes described in Section 2.2. For each sensor, the reference SNR is obtained on its desert calibration scene and then radiance-normalised under the shot-noise scaling assumption (SNR ∝ √L) to suppress brightness differences between calibration targets [17]. PRISMA and Tanager-1 curves are scaled to the EnMAP reference radiance regime (EnMAP Agadez calibration scene), thereby isolating cross-sensor differences in the effective instrument noise floor relevant to HyGAS uncertainty propagation [16], [17].

### *B. Striping diagnostics and implications for matched-filter retrieval (Fig. 3)*

The full striping assessment in the extended paper [17] shows that across-track non-uniformity differs systematically among PRISMA, EnMAP, and Tanager-1 in the methane SWIR windows, with direct consequences for background stability in MF enhancement products. For the sake of brevity, Fig. 3 (placed at the end of the document for better readability) reports only the ratio-based striping amplitude (adjacent-band ratio diagnostic), which is designed to suppress residual surface/illumination structure and highlight detector-pattern artefacts.

Across the methane windows, the ratio diagnostic indicates that EnMAP exhibits the lowest residual striping, consistent with its strong inter-column radiometric coherence; PRISMA shows larger column-scale non-uniformity, particularly in the SWIR; and Tanager-1 displays the most pronounced column-dependent variability, despite operating in a favourable radiance-normalised SNR regime (Fig. 1). This combination—high median SNR but stronger inter-column variability—explains why scene-wide background statistics can yield enhancement maps with visibly structured false positives for Tanager-1 (and, to a lesser extent, PRISMA), while column-wise CMF (CW-CMF) better absorbs column-correlated variability into the background model, reducing striping leakage into ΔX and stabilising downstream plume segmentation and IME/flux estimates. The complete comparison across all striping metrics and methane windows is reported in [17] (Fig. 15), and the operational impact of striping suppression is illustrated by the Buenos Aires revisit case study discussed in [17].

### *C. Buenos Aires landfill case study (CW-CMF) (Fig. 2)*

Figure 2 reports the Buenos Aires landfill super-emitter revisit pair (Section 2.2), processed with a consistent CW-CMF configuration and the same scale-aware segmentation framework [16], [17]. To remain within IGARSS space, only CW-CMF enhancement maps (and plume masks) are shown, highlighting improved background stability over a heterogeneous urban–industrial context relative to scene-wide statistics. Under CW-CMF, the plume-integrated results (from the extended manuscript) are: Tanager-1 IME = 30,590.94 ± 101.29 kg, Q = 31.53 ± 8.21 t $h^{-1}$, and EnMAP IME = 98,858.52 ± 481.67 kg, Q = 74.02 ± 16.29 t $h^{-1}$ [17]. Due to the fact that the acquisitions are non-simultaneous, these values are not interpreted as a direct inter-sensor disagreement in emission rate; rather, they provide a controlled demonstration—over the same facility type and a difficult background—of how sensor radiometry and MF configuration impact the stability of ΔX and the robustness of uncertainty-aware IME/flux estimates within a unified pipeline [16], [17].

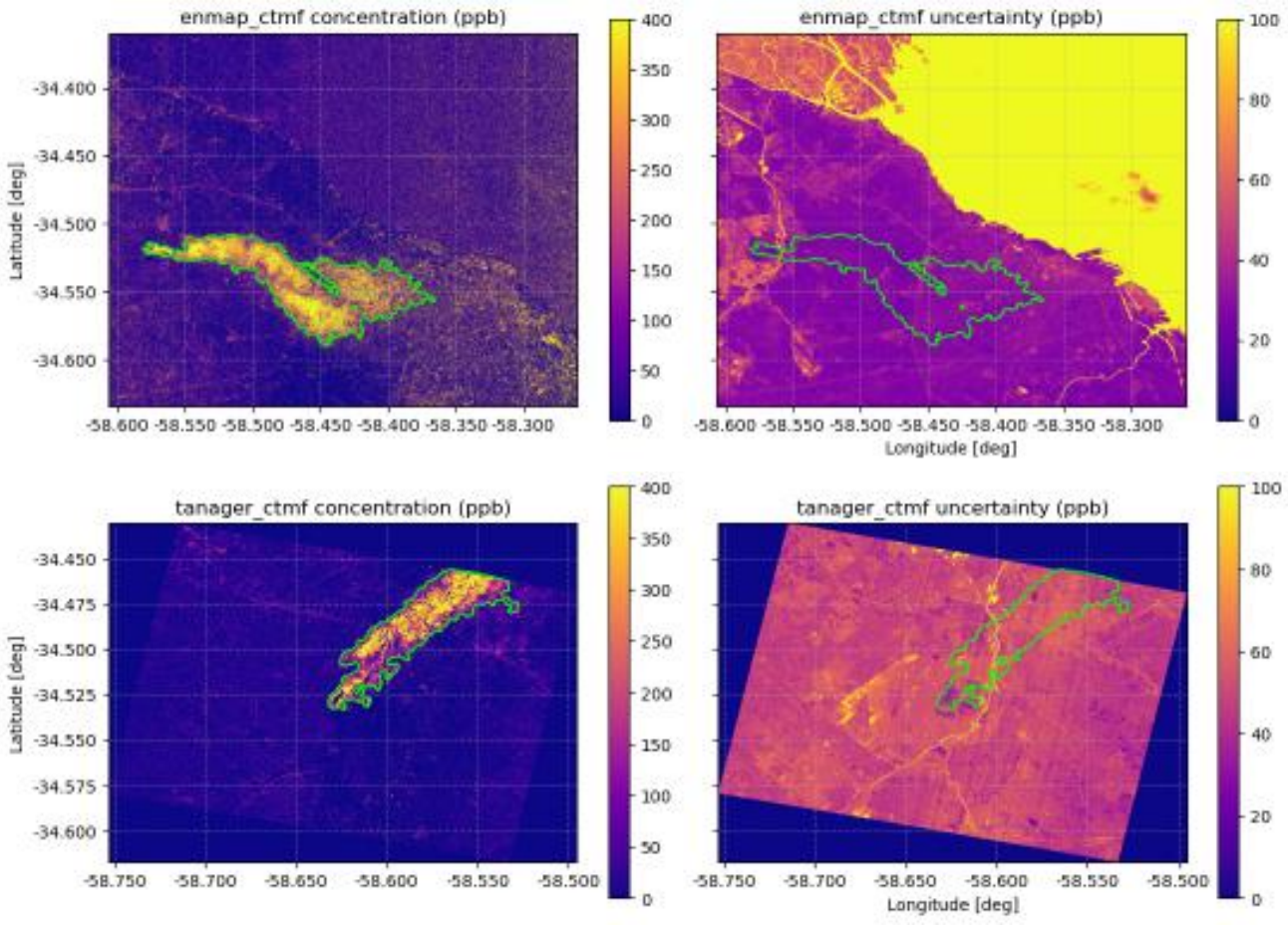

**Fig. 2.** Buenos Aires landfill super-emitter (Complejo Ambiental Norte III): CW-CMF methane enhancement ΔX (ppm·m) and segmented plume mask

for EnMAP L1B_DT0000064332_014_20240309T142431Z (9 March 2024) and Tanager-1 20250626_144419_75_4001 (26 June 2025). IME/flux values reported in [17] and summarised in Section 4.3.

Although HyGAS supports a comprehensive intercomparison workflow and can directly leverage near-simultaneous multi-sensor matchups where available [16], [17], in the present Tanager-1 integration we did not find near-simultaneous Tanager-1 acquisitions overlapping EnMAP or PRISMA over the same sources within the Planet Core imagery collection. Consequently, the Buenos Aires pair is non-simultaneous and IME/flux differences are interpreted as a stability assessment over a difficult, heterogeneous background rather than as absolute inter-sensor disagreement [17]. Future work will target near-simultaneous matchups as they become available and will include evaluation of Planet's distributed $CH_4$ products where provided [11]–[13], [16], [17].

## V. Conclusions

This paper presents a technical integration and assessment of Tanager-1 methane retrieval within the HyGAS unified multi-sensor pipeline previously applied to PRISMA and EnMAP [7]–[9], [14]–[17]. Calibration-scene diagnostics show that Tanager-1 can operate in a favourable SNR regime in the methane window, but with stronger inter-column variability that motivates column-aware covariance estimation [11]–[13], [17]. The Buenos Aires landfill revisit case study demonstrates CW-CMF operational outputs and uncertainty-aware IME/flux quantification over a difficult background, emphasizing sensitivity and robustness rather than absolute accuracy [4]–[6], [16], [17].

The processing described here is implemented in HyGAS [16]. The repository is scheduled for open-source release upon publication of the accompanying journal article [17]. In the interim, a frozen version can be shared with reviewers upon request for reproducibility.

## Acknowledgment

This research was supported by the Italian Space Agency ASI within the CLEAR-UP project (Contract/Agreement n. 2022-16-U.0; CUP n. F83C22000780005). We acknowledge Planet for providing access to Tanager-1 Core Imagery data used in this study. We also gratefully acknowledge the German Aerospace Center (DLR) for providing EnMAP data.

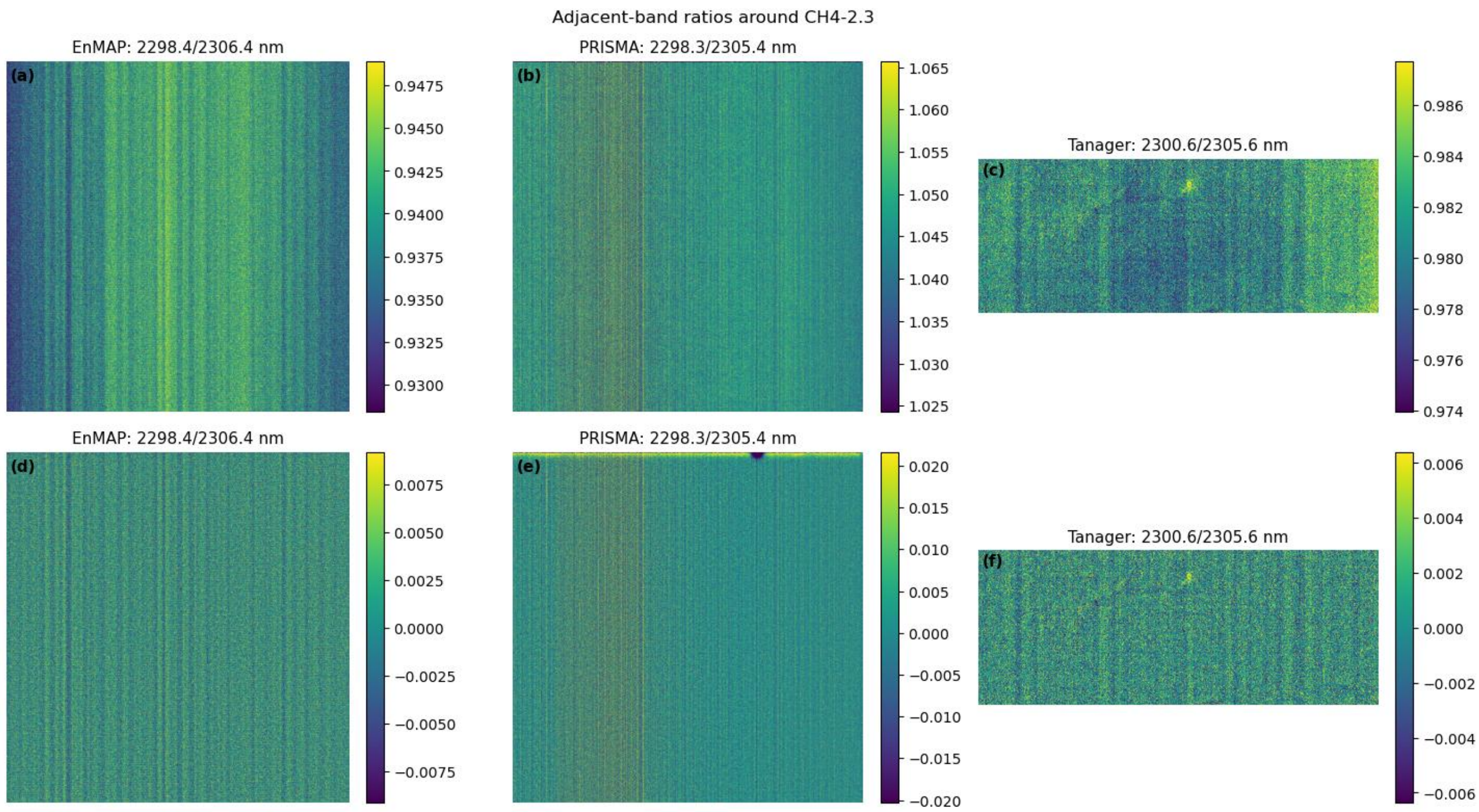

**Fig. 3**. Ratio-based striping amplitude in the methane SWIR window(s) for PRISMA, EnMAP, and Tanager-1, computed from homogeneous calibration scenes (Section II.B). The metric is derived from an adjacent-band ratio image to suppress surface brightness structure while preserving detector-pattern artefacts coherent across nearby wavelengths, then summarised with robust statistics over the methane window. Lower values indicate cleaner across-track uniformity. This figure corresponds to the ratio-metric panel of the full striping analysis reported in the extended paper [17] (see Fig. 15 therein).